\documentclass[12pt]{article}
\usepackage{graphicx}

\usepackage[latin1]{inputenc}
\usepackage{shortvrb}
\usepackage{epsfig}
\usepackage{amssymb}
\usepackage{mathrsfs}
\usepackage{amsbsy}
\usepackage{graphics}
\usepackage{epsf}
\usepackage{amsmath}
\usepackage{amssymb}
\usepackage[usenames]{color}
\usepackage{pstricks}

\usepackage{indentfirst}

\newcommand{\beq}{\begin{equation}}
\newcommand{\eeq}{\vspace{0cm} \end{equation}}
\newcommand{\beqq}{\setlength\arraycolsep{2pt}\begin{eqnarray}}
\newcommand{\eeqq}{\vspace{0cm} \end{eqnarray}}

\newcommand{\hsp}{\hspace{1cm}}

\begin{document}
\begin{center}
\large {\bf ON THE CHEMICAL POTENTIAL OF DARK ENERGY}
\end{center}
\begin{center}
{\bf S. H. Pereira}
\end{center}
\vspace{0.05cm}
\begin{center}
{\bf  Departamento de Astronomia, Universidade de S\~ao Paulo \\ Rua do Mat\~ao, 1226 - 05508-900, S\~ao Paulo, SP, Brazil}\\
\noindent spereira@astro.iag.usp.br
\end{center}

\begin{abstract}
It is widely assumed that the observed universe is accelerating due to the existence of a new fluid component called dark energy. In this article, the thermodynamics consequences of a nonzero chemical potential on the dark energy component is discussed with special emphasis to the phantom fluid case.  It is found that if the dark energy fluid is endowed with a negative chemical potential, the phantom field hypothesis becomes thermodynamically consistent with no need of negative temperatures as recently assumed in the literature.
\end{abstract}




\section{Introduction}

The  current idea of an accelerating Universe driven by dark energy
is based on a  large convergence of independent
observational results, and its explanation constitutes one of the greatest
challenges for our current understanding of fundamental physics \cite{SN,CMB}. The origin and the nature of dark energy
is still a mystery, however, there is no doubt that its existence is
beyond the domain of the standard model of particle physics \cite{review}.

Among a number of possibilities to describe this
dark energy component, the simplest and most theoretically
appealing way is by means of a cosmological constant $\Lambda$,
which acts on the Einstein field equations as an isotropic and
homogeneous source with a constant equation of state parameter $p/\rho=-1$.
On the other hand, although cosmological scenarios with a $\Lambda$ term
might explain most of the current astronomical observations, from the
theoretical viewpoint they are plagued with some fundamental problems 
thereby stimulating the search for alternative dark energy models 
driven by different candidates \cite{model,list1}.  

In 1995, Lima and collaborators \cite{limamaia,LA04} analyzed several thermodynamic and statistical properties of a dark energy fluid. They  assumed a dark energy fluid component phenomenologically described by an equation of state $p=\omega \rho$ with null chemical potential. Later on, Lima and Alcaniz (and independently Brevik et al. \cite{B}) stressed that their theoretical  thermodynamic treatment ruled out the case of
phantom energy because the comoving entropy of a dark component with $\omega < -1$ is negative (see Reference \cite{REF} for other interesting discussions of phantom fluids).  However,  thermodynamic arguments in favor of the phantom hypothesis
were put forward by Gonz\'alez-D\'{i}az and Sig\"uenza \cite{gonzalez}. They claimed that the temperature of a phantomlike
fluid is always negative in order to keep its  entropy positive
definite (as statistically required).

In this work we reanalyze the thermodynamics properties of an expanding universe  filled with a dark
energy fluid endowed with a non-zero chemical potential. As we shall see, the main effect of a chemical potential is that the phantom scenario becomes thermodynamically consistent with no need to assume negative temperatures.

\section{Thermodynamics properties of the dark energy}


Let us now consider that the Universe is described by the
homogeneous and isotropic Friedmann-Robertson-Walker (FRW) geometry
($c=1$)
\begin{equation}
 ds^2 = dt^2 - a^{2}(t) \left(\frac{dr^2}{1 - \kappa r^{2}}
  + r^2 d \theta^2 +
 r^2 sin^{2}\theta d\phi^2\right),
\end{equation}
where $\kappa = 0,\pm 1$ is the curvature parameter and $a(t)$ is
the scale factor. In what follows we consider that the matter
content is a fluid
described  by the equation of state \beq p=\omega \rho\,,\label{eqstate} \eeq
where $p$ is the pressure, $\rho$ is the energy density and $\omega$
a constant parameter which may be positive (white energy) and
negative (dark energy).  The cases $\omega = 1/3$, $1$,  and $-1$
characterizes, respectively, the blackbody radiation, a stiff-fluid
and the vacuum state while $\omega < -1$ stands to a phantomlike
behavior.

Following standard lines, the equilibrium thermodynamic states of a
relativistic simple fluid are characterized by an energy momentum
tensor $T^{\alpha \beta}$, a particle current $N^{\alpha }$ and an
entropy current $S^{\alpha}$ which satisfy  the following relations
\begin{equation}
T^{\alpha \beta}=(\rho + p)u^{\alpha} u^{\beta} - pg^{\alpha \beta},
\quad T^{\alpha \beta};_{\beta}=0 , \label{eq:TAB}
\end{equation}
\begin{equation} \label{eq:NA}
N^{\alpha}=nu^{\alpha}, \quad  N^{\alpha};_{\alpha}=0 ,
\end{equation}
\begin{equation} \label{eq:SA}
S^{\alpha}=s u^{\alpha}, \quad  S^{\alpha};_{\alpha}=0 ,
\end{equation}
where ($;$) means covariant derivative, $n$ is the particle number
density, and $s$ is the entropy density. In the FRW background, the
above conservation laws can be rewritten as (a dot means comoving
time derivative)
 \begin{equation}
 \dot{\rho} + 3 (1 + \omega)\rho \frac{\dot a}{a}=0,\hsp
 \dot{n} + 3n\frac{\dot a}{a}=0,\hsp \dot{s} + 3s\frac{\dot a}{a}=0,
\end{equation}
whose solutions can be written as:
\begin{equation}
\rho=\rho_0 \left(\frac{a_0}{a} \right)^{3(1 + \omega)}, \hsp n=n_0
\left(\frac{a_0}{a}\right)^{3}, \hsp s=s_0
\left(\frac{a_0}{a}\right)^{3},\label{eqSol}
\end{equation}
where the positive constants $\rho_0$, $n_0$, $s_0$ and $a_0$ are
the present day values of the corresponding quantities (hereafter
present day quantities will be labeled by the index ``0"). On the other
hand, the quantities $p$, $\rho$, $n$ and $s$ are related to the
temperature $T$ by the Gibbs law
\begin{equation} \label{eq:GIBBS}
nTd(\frac{s}{n})= d\rho - {\rho + p \over n}dn,
\end{equation}
and from Gibbs-Duhem relation \cite{callen} there are only two
independent thermodynamic variables, say, $n$ and $T$. Now, by
assuming that $\rho=\rho(T,n)$ and $p=p(T,n)$ and combining  the
thermodynamic identity \cite{weinb}
\begin{equation}
T \biggl({\partial p \over \partial T}\biggr)_{n}=\rho + p - n
\biggl({\partial \rho \over \partial n}\biggr)_{T},
\end{equation}
with the conservation laws as given by (6), one may show that the
temperature satisfies
\begin{equation} \label{eq:EVOLT}
{\dot T \over T} = \biggl({\partial p \over \partial
\rho}\biggr)_{n} {\dot n \over n} = -3\omega \frac{\dot a}{a}.
\end{equation}
Therefore, assuming that $\omega \neq 0$ we obtain
\begin{equation} \label{eq:TV}
n = n_0 \left(\frac{T}{T_0} \right)^{1 \over  \omega}  \quad
\Leftrightarrow \quad T=T_0 \left(\frac{a}{a_0} \right)^{-3\omega}.
\end{equation}
The temperatures appearing in the above expressions are positive
regardless of the value of $\omega$.  In particular, in the
radiation case ($\omega = 1/3$), one finds $aT=a_0T_0$ as should be
expected. As compared to this case, the unique difference is that
the dark energy fluid (even in the phantom regime) becomes hotter in
the course of the cosmological adiabatic expansion since its
equation of state parameter is a negative quantity. A physical
explanation for this behavior is that thermodynamic work is being
done on the system \cite{limamaia,LA04}.

It should  be stressed that the derivation of the temperature
evolution law presented here is fully independent of the entropy
function, as well as of the chemical potential $\mu$. The above
expressions also imply that for  any comoving volume of the fluid,
the product $T^{1 \over  \omega} V$  remains constant in the course
of expansion and must also characterize the equilibrium states
(adiabatic expansion) regardless of the value of $\mu$. Further, by
inserting the temperature law into the energy conservation law
(\ref{eqSol}), one obtains the energy density as function of the
temperature
\begin{equation}
\rho= \rho_0 \left(\frac{T}{T_0} \right)^{\frac{1 +
\omega}{\omega}}.
\end{equation}
Now, in order to determine the chemical potential and its influence
on the thermodynamic of dark energy,  we consider the Euler
relation\cite{callen}
\beq Ts={p+\rho}-{\mu n},\,\label{entropyd}
\eeq
where $\mu$ in general can also be a function of $T$ and $n$
\cite{FUG,degroot}. By combining the above expression with equations
(\ref{eqstate}), (\ref{eqSol})  and (\ref{eq:TV}) we obtain:
\beq \mu=\mu_0\left(\frac{a}{a_0} \right)^{-3\omega} =
\mu_0\left(\frac{T}{T_0} \right), \eeq where \beq\label{mu0}
\mu_0={1\over n_0}[(1+\omega)\rho_0-T_0 s_0].
\eeq
This straightforward thermodynamic result has some interesting
consequences.  Remember that all the present day quantities labeled by the index ``0''  
are positives. In principle, the chemical potential may be either
positive or negative, and it also  depends on the values of the
$\omega$-parameter. In particular,  $\mu$ is always negative ($\mu_0
< 0$) in the case of phantom  energy, and becomes even more
negative in the course of time ($T$ grows with the scale factor
during the cosmic evolution). It is also known that $\mu$ is zero in
the case of photons ($\omega=1/3$) because they are their own
antiparticles.  In this case, (\ref{mu0}) yields
correctly that  $s_0 T_0=(4/3)\rho_0$ as should be expected. In
general, if $\mu=0$, necessarily  the  relation $s_0 T_0= (1 +
\omega)\rho_0$ must be obeyed, which is just the present day
expression of $sT=(1 + \omega)\rho$ as required by (\ref{entropyd}).

At this point, the fundamental question is: How the chemical
potential modifies the entropy constraints
derived in the previous papers \cite{limamaia,LA04}?

In order to show that we compute explicitly the entropy of dark
energy for a comoving volume $V$. As remarked before, the entropy
function should scale as $S \propto T^{1 \over \omega}V$. Actually,

\beq S(T,V)\equiv sV = \left[\frac{(1+\omega)\rho_0-\mu_0
n_0}{T_0}\right] \bigg({T\over
T_0}\bigg)^{1/\omega}V=s_0V_0,\label{entropy} 
\eeq 
which remains
constant as expected (see discussion below Eq.(\ref{eq:TV})).
However, in order to keep the entropy $S\geq 0$ (as statistically
required), the following constraint must be satisfied: \beq \omega
\geq \omega_{min}=-1 + {\mu_0 n_0 \over \rho_0}\,,\label{accord}
\eeq which introduces a minimal value to the $\omega$-parameter,
below which the entropy becomes negative. This is a remarkable
expression and its consequences are apparent. For instance, consider
that $\mu_0=0$ (no chemical potential). In this case,  the smallest
value of the $\omega$-parameter is $\omega_{min}=-1$ and the
previous analysis by Lima and Alcaniz \cite{LA04} is fully
recovered, that is, the phantom domain ($\omega < -1$) is
thermodynamically forbidden. However, for a negative chemical
potential, the phantomlike regime becomes thermodynamically allowed
thereby recovering the hypothesis of phantom energy without
appealing to negative temperature as proposed in the literature \cite{gonzalez}.  Note also that for a positive chemical potential
not even a cosmological constant ($\omega = -1$) is possible. In
Figure 1, we summarize the basic thermodynamic results.

\begin{figure}[htb]
\begin{center}
\epsfig{file=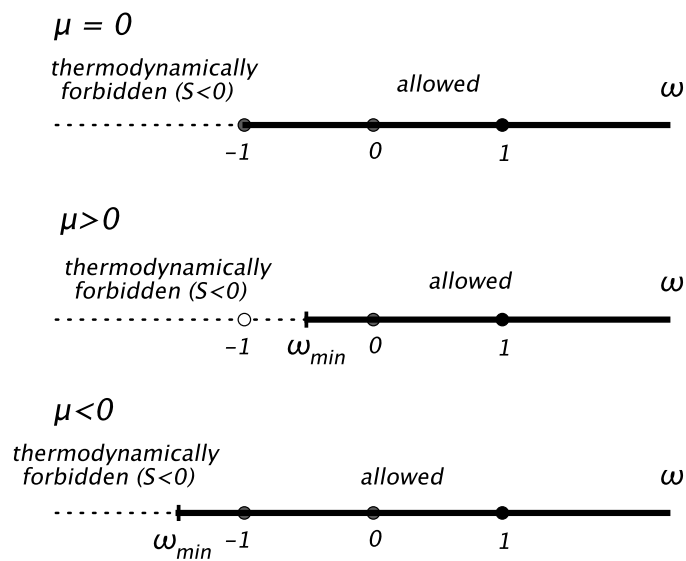, scale=0.8}
\end{center}
\caption{The allowed intervals of $\omega$ values (heavy lines) and
forbidden (dashed lines) for null, positive and negative chemical
potentials. Note that a large portion of the dark branch $\omega <
0$ is always  thermodynamically permitted. However,  for $\mu \geq
0$, the phantomlike behavior ($\omega<-1$) is thermodynamically
forbidden. }\label{figt}
\end{figure}

\section{Concluding Remarks}

In this paper we have investigated the thermodynamics properties of a dark energy fluid with equation of state, $p=\omega
\rho$, by assuming that its chemical potential is different from
zero.

In Figure 1, we summarize the main results of the present analysis
including a chemical potential to the dark energy fluid.  As discussed in previous section, the
regions with $S < 0$ are  thermodynamically forbidden. Note also
that many dark energy fluids satisfy the combined constraints
regardless of the $\mu$ sign, that is, a large interval of  negative
$\omega$ values is allowed from thermodynamic 
considerations.  However, a phantom like behavior ($\omega < -1$) is
permitted only for $\mu < 0$. It should be stressed that for $\mu=0$ one finds  $\omega_{min}=-1$
(see Eq. (\ref{accord})) in accordance to the results previously
derived by Lima and Alcaniz \cite{LA04}. The present analysis with
$\mu \neq 0$  also opens  the possibility for an equation of state parameter $\omega <
-1$, thereby recovering the idea of a phantom dark energy with no need of negative temperatures.  Therefore, as far as we known, the inclusion of a negative chemical potential to the dark energy fluid is the only way to save the phantom hypothesis without to violate basic thermodynamic properties.

\section*{Acknowledgments}

SHP is supported by CNPq No. 150920/2007-5 (Brazilian Research Agency).


\begin{thebibliography}{99}

\bibitem{SN} A. G. Riess {\it et al.}, Astron. J. {\bf{116}}, 1009 (1998); 
S. Perlmutter {\it et al.}, Astrophys.  J. {\bf{517}}, 565 (1999);  
P. Astier et al., Astron. Astrophys. {\bf 447}, 31 (2006); A. G. Riess et al., Astro. J. {\bf 659}, 98 (2007).

\bibitem{CMB} D. N. Spergel et al. Astrophys. J. Suppl. Ser. {\bf 170}, 377 (2007);  
S. W. Allen et. al., [arXiV:0706.0033] (2007).

\bibitem{review} T. Padmanabhan, Phys. Rept. {\bf 380}, 235 (2003); 
P. J. E. Peebles and  B. Ratra,  Rev. Mod. Phys. {\bf 75}, 559 (2003);
J. A. S. Lima, Braz. Jour. Phys. {\bf 34}, 194 (2004),
[astro-ph/0402109]; V. Sahni and A. Starobinsky, Int. J. Mod. Phys. D {\bf 15},
2105 (2006).

\bibitem{model} S. Weinberg, Rev. Mod. Phys. {\bf 61}, 1 (1989); D. Pav\'{o}n, Phys.
Rev. D {\bf 43}, 375 (1991); J. C. Carvalho, J. A. S. Lima and I.
Waga, Phys. Rev. D {\bf{46}}, 2404 (1992); J. A. S. Lima and J. M.
F. Maia, Phys. Rev D {\bf 49}, 5597 (1994); J. A. S. Lima and J. C. Carvalho, Gen. Rel. Grav. {\bf 26}, 909 (1994); 
J. A. S. Lima and M. Trodden, Phys. Rev. D {\bf 53}, 4280 (1996),[astro-ph/9508049]; J.
A. S. Lima, Phys. Rev. D {\bf 54}, 2571 (1996); J. M. Overduin and F. I. Cooperstock, Phys. Rev. D {\bf 58}, 043506 (1998);  J. V. Cunha and R. C. Santos, Int. J. Mod. Phys. D {\bf 13}, 1321 (2004), [astro-ph/0402169];  J. F. Jesus, [astro-ph/0603142] (2006);  J. F. Jesus {\it et al.}, [arXiv:0806.1366].   
 
\bibitem{list1} T. Padmanabhan and T. R. Choudhury,
Mon. Not. Roy. Astron. Soc. {\bf 344}, 823 (2003); P. T. Silva and
O. Bertolami, Astrophys.  J.  {\bf 599}, 829 (2003); J. A. S. Lima,
J. V. Cunha and J. S. Alcaniz, Phys. Rev. D {\bf 68}, 023510 (2003);  Z. H. Zhu and M. K. Fujimoto, Astrophys. J. {\bf 585},
52 (2003);  S. Nesseris and L. Perivolaropoulos, Phys. Rev. D
{\bf{70}}, 043531 (2004);  Y. Wang and M. Tegmark, Phys. Rev. Lett.
{\bf{92}}, 241302 (2004); T. R. Choudhury and T. Padmanabhan,
Astron. Astrophys. {\bf 429} 807 (2005); F. C. Carvalho {\it et al.}, Phys. Rev. Lett. {\bf 97}, 081301 (2006), [astro-ph/0608439]; {\bf ibdem}, [arXiv:0704.3043] (2007); J. V. Cunha, L. Marassi and R. C. Santos, Int. J. Mod. Phys. D {\bf 16}, 403 (2007); R. C. Santos and J. A. S. Lima, Phys. Rev. D {\bf 77}, 023519 (2008), [arXiv:0803.1865].  





\bibitem{limamaia} J. A. S. Lima and A. Maia Jr., Phys. Rev. D {\bf 52},  5628 (1995); {\bf ibdem},  Int. J. Theor. Phys, {\bf 34}, 9  (1995), [gr-qc/9505052]; J. A. S. Lima and J. Santos, Int. J. Theor. Phys. {\bf 34}, 143 (1995); J. A. E. Carrillo, J. A. S.
Lima, A. Maia Jr.,  Int. J. Theor. Phys. {\bf 35}, 2013 (1996).
[hep-th/9906016].

\bibitem{LA04} J. A. S. Lima and J. S. Alcaniz, Phys. Lett. B {\bf 600}, 191 (2004), [astro-ph/0402265].

\bibitem{B} I. Brevik, S. Nojiri, S. D. Odintsov and  Luciano Vanzo, Phys. Rev. D{\bf 70}, 043520 (2004).

\bibitem{REF} H. M. Sadjadi, Phys. Rev. D {\bf 73}, 063525 (2006), M. R. Setare and S. Shafei, JCAP 0609, 011 (2006);
B. Wang, Y. Gong, and E. Abdalla, Phys. Rev. D {\bf 74}, 083520 (2006); F.C. Santos, M. L. Bedran, and V. Soares, Phys. Lett. B {\bf 636}, 86 (2006); R. C. Santos and J. A. S. Lima, [astro-ph/0609129];  Y. Gong, B. Wang, and A. Wang, Phys. Rev. D {\bf 75}, 123516 (2007); N. Bilic, Fortschr. Phys. {\bf 56}, 363 (2008), [arXiv:0806.0642].  

\bibitem{gonzalez}
P. F. Gonz\'alez-D\'{i}az and  C. L. Sig\"{u}enza, Nucl. Phys. B {\bf
697},363 (2004); Phys. Lett. B {\bf 589}, 78 (2004).

\bibitem{callen} H. B. Callen, {\it Thermodynamics and an Introduction to Thermostatistics}, 2nd Edition, Wiley, New York (1985).

\bibitem{weinb} S. Weinberg, Astrophys. J. {\bf 168}, 175 (1971); R.
Silva, J. A. S. Lima and M. O. Calv\~ao, Gen. Rel. Grav. {\bf 34},
865 (2002), [gr-qc/0201048].


\bibitem{FUG} In the case of a relativistic ideal gas of massive particles, for instance,
the so-called  fugacity reads, $exp(\mu/k_B T) = n(2\pi\bar
h)^{3}/4\pi m^{2}ck_B K_2(mc^{2}/k_B T)$, where $m$ is the particles
mass and $K_2$ is the modified  Bessel of second kind.

\bibitem{degroot}S. R. de Groot, W. A. van Leewen and Ch. G. van
Weert, {\it Relativistic Kinetic Theory}, North Holland, Amsterdam
(1980).

\end{thebibliography}
\end{document}